\documentclass[pdflatex,sn-mathphys]{sn-jnl}

\jyear{2023}%

\theoremstyle{thmstyleone}%
\theoremstyle{thmstyletwo}%

\theoremstyle{thmstylethree}%
\makeatletter \def\ps@pprintTitle{  \let\@oddhead\@empty  \let\@evenhead\@empty  \def\@oddfoot{\hfill\thepage}  \def\@evenfoot{\thepage\hfill}} \makeatother
\usepackage{caption}
\usepackage{subcaption}
\usepackage{hyperref}
\usepackage{url}
\usepackage{natbib}

\begin{document}


\title [  ]{A Hybrid Feature Fusion Deep Learning Framework for Leukemia Cancer Detection in Microscopic Blood Sample Using Gated Recurrent Unit and Uncertainty Quantification }


\author[1]{\fnm{Maksuda} \sur{Akter}}\email{maksudaoni6@gmail.com}

\author[2]{\fnm{Rabea} \sur{Khatun}}\email{rabea@cse.green.edu.bd}

\author*[1]{\fnm{Md Manowarul} \sur{Islam}}\email{manowar@cse.jnu.ac.bd}


\affil[1]{\orgdiv{Department of Computer Science and Engineering}, \orgname{Jagannath University}, \orgaddress{\city{Dhaka}, \country{Bangladesh}}}

\affil[2]{\orgdiv{Department of Computer Science and Engineering}, \orgname{Green University of Bangladesh}, \orgaddress{\city{Dhaka}, \country{Bangladesh}}}


\abstract{}
Acute lymphoblastic leukemia (ALL) is the most malignant form of
leukemia. It is considered the most prevalent kind of cancer in both
adults and children. Leukemia is typically identified by analyzing blood
and bone marrow smears under a microscope. Advanced cytochemical
testing can also be performed to confirm and categorize leukemia. How-
ever, these approaches are expensive, time-consuming, and dependent
on the skill and knowledge of the relevant specialists. In recent decades,
deep learning with Convolutional Neural Networks (CNN) has generated
sophisticated methods for image classification by analyzing microscopic
smear images to detect the presence of leukemic cells. These techniques
are easy to use, quick, affordable, and free from expert bias. Nevertheless,
the majority of these approaches are unable to quantify the uncertainty
in their results, which can have devastating consequences. This research
implemented hybrid models (InceptionV3-GRU, EfficientNetB3-GRU,
and MobileNetV2-GRU) for the ALL detection and classification. Following this, to identify an optimal set of hyperparameters and enhance the model's performance, bayesian optimization is utilized. Then,
we use Deep Ensemble uncertainty quantification approach to deal with
uncertainty during leukemia image classification. The outcomes of three different hybrid deep learning models were then aggregated at the score level using the sum rule after they had been trained on the two publicly available leukaemia patient blood sample datasets, ALL-IDB1 and ALL-IDB2. Herein, parallel architecture was considered,
which offers a high degree of confidence in discriminating between ALL and
not ALL. The proposed system has managed to correctly and accurately
diagnose the leukemia patients, respectively, with a detection accuracy
rate of 100\% for the ALL-IDB1 dataset, 98.07\% for the ALL-IDB12 dataset, and
98.64\% for the combined dataset using our proposed method.


\maketitle

\section{Introduction}
\label{intro}
Leukemia is the most severe type of blood cancer across all age groups. Excessive and immature blood cell proliferation is the origin of this anomalous condition\cite{oviedo2021journal}, which can harm bone marrow, the immune system, and red blood cells\cite{ghaderzadeh2021machine,hegde2020automated}.
Blood is a vital component of the human body, consisting of 45\% red blood cells and 55\% plasma. Depending on the blood's size, composition, texture, color, and shape, there are three primary blood components: thromocytes (platelets), leukocytes which is commonly known as white blood cells (WBC), and erythrocytes which is known as red blood cells (RBC).
RBC rate ranges from 4,000,000 to 6,000,000 per microliter of blood which carries oxygen throughout the bodywbc. WBC provide immunity and resistance and protect our body against infections; their typical concentration in the body is between 4,000 and 11,000 cells per microliter. Blood clotting is carried out by platelets, which have a density of 150,000 to 450,000 per microliter of blood. Therefore, alterations in any of the fundamental blood constituents will lead to health issues for an individual.

Leukemia is a fatal cancer that results in an abnormal
increase in immature WBC in the bone marrow and blood. A high WBC volume encompasses both RBCs and platelets, resulting in a low level of body immunity. There are several possible causes of leukemia, including radiation, familial history, and environmental pollutants\cite{henry2006clinical}. However, the true cause of the disease is unknown\cite{anilkumar2021automated}. Depending on how rapidly it grows, leukemia can be characterized as either acute or chronic. WBC that are contaminated with acute leukemia cannot function or behave normally like normal WBC. However, in chronic leukemia, they may behave normally\cite{ahmed2019identification}. Myeloid or lymphoid is the classification given
to leukemia based on the original cell line. Therefore, based on the disease's course, severity, and impact, specialists divide it into four categories: Acute Lymphoblastic Leukemia (ALL), Acute Myeloblastic Leukemia, Chronic Lymphocytic Leukemia and Chronic Myeloblastic leukemia\cite{mohapatra2010image}\cite{patel2015automated}. Among these, ALL accounts for 70\% of all instances of leukemia and is also the most deadly\cite{sawyers1991leukemia}\cite{abunadi2022multi}. 

To reduce medical risks and choose the best course of treatment, it's critical to determine whether leukemia exists and what kind it is. Hematologists analyze bone marrow or blood smears under a microscope in clinical laboratories to detect ALL and its subtypes\cite{rehman2018classification}. The number of distinct blood cells and their morphological characteristics are usually used to classify leukemia cases. Generally, these leukemic cells are easy to identify due to their dark-purple hue, but because of the variances in pattern and texture, assessment and further processing become highly intricate. The lymphocytes, or lymphoblasts, in ALL individuals have a very thin, homogenous border. They also have spherical particles called nucleoli inside the nucleus and microscopic pockets in the cytoplasm called vacuoles. As the disease progresses, it becomes increasingly clear that the prescribed medication could cause an early death if it is ignored\cite{bukhari2022deep}. 

Cancer, a fatal disease that is primarily caused by metabolic problems and inherited conditions, is one of the leading causes of death. According to the World Health Organization (WHO), cancer ranks as the primary cause of death before the age of 70 in 112 out of 183 nations, and as the third or fourth in 23 more\cite{talukder2023efficient}\cite{khatun2023cancer}. Children under the age of five have the highest chance of developing ALL. The risk decreases gradually until the mid-20s and then starts to rise slowly once more around age 50\cite{namayandeh2020global}. Data from WHO 2020 shows that there were around 474,519 cases of leukemia worldwide, which resulted in 311,594 fatalities. According to the American Cancer Society's estimates, there were around 6,540 new cases of ALL and 1,390 deaths from the disease in the US in 2023. Ref. \cite{dwivedi2018artificial} reports that 52,995 people died and around 153,315 people worldwide experienced ALL in 2019\cite{bukhari2022deep}. With early diagnosis and treatment, patients chances of survival can rise. As a result, early detection and successful treatment of ALL are crucial\cite{haworth1981routine}\cite{bain2005diagnosis}. In order to validate and distinguish common subtypes of leukemia, additional procedures like flow cytometry, chromosomal analysis, and complex cytochemical testing could be needed. These intricate procedures are time-consuming, expensive, and dependent on the medical team's knowledge and proficiency in sample collection, preparation, and testing. For a routine examination, the complex laboratory tests are also not that simple\cite{putzu2014leucocyte}\cite{agaian2014automated}. In order to overcome the shortcomings associated with manual screening, hematologists need to develop an automated approach that can identify or categorize malignant lymphocytes\cite{husham2016automated}\cite{iqbal2018computer}.

In comparison to several intricate clinical procedures, automated image processing-based methods are simple, quick, and less expensive while yet producing a diagnosis that is accurate and timely. Issues with manual diagnostics can be fixed by Machine Learning (ML) and Deep Learning (DL) techniques. DL algorithms are able to automatically extract features from raw data. Traditional machine learning methods can be outperformed by deep learning models, which can attain state-of-the-art performance on picture classification tasks. In this study, DL models and hybrid DL models were used to examine the two datasets,\texttt{ ALL\_IDB1} and \texttt{ ALL\_IDB2}, for leukemia diagnosis.

Three DL models, namely the MobileNetV2 GRU (MobileNetV2-GRU), the InceptionV3 gated recurrent unit (InceptionV3-GRU), and the EfficientNetb3 gated recurrent unit (EfficientNetb3-GRU), are used in this study to carry out the feature extraction and classification tasks for ALL diagnosis in microscopic images. To avoid making predictions with too much confidence, uncertainty qunatification (UQ) is utilized. Ultimately, the final decision is reached by combining the output produced by these three deep learning models. The primary driving force for the creation of the suggested system is to create an efficient and effective system that can be
developed in a real-world clinical situation for fast
diagnosis and treatment. The main contributions of the suggested work for automated leukemia detection are outlined below:

\begin{itemize} 
    \item  For significantly high accuracy classification of medical imaging data, we put forth a unique hybrid DL model. wherein the model integrates three hybrid DL models (MobileNetV2-GRU, InceptionV3-GRU, and EfficientNetb3-GRU) that precisely identify ALL and not ALL while also automatically extracting features from the dataset.
    \item To avoid overconfidence in the diagnosis of the disease, we employed well-known uncertainty quantification approach, such as Deep Ensemble (DE), to increase the classification stages confidence level. 
    \item By implementing a Bayesian optimization technique, we optimized hyperparameters of deep learning models to ensure that it performs at its best.
    \item The final decision in the suggested system is obtained by combining the
results generated from three different hybrid DL models trained
on \texttt{ ALL\_IDB1} and \texttt{ ALL\_IDB2} datasets, using the
sum rule at the score-leve.
\end{itemize}

The rest of the paper is organized as follows: a review of relevant
Previous studies are presented in Section 2. 
The proposed approach and the classification algorithms used in the proposal are explained in Section 3. The experiment results are presented in Sections 4 and 5, along with a discussion. Section 6 wraps up our study and discusses its shortcomings and future efforts.

\section{Related works }
\label{Related}
An overview of relevant studies that use Deep Learning (DL) and Machine Learning(ML) for automated 
Leukemia detection is presented in Tables \ref{tab:rw}.

K.K. Anilkumar \textit{\textit{et al.}} \cite{anilkumar2021automated} proposed a methodology that made use of the Directed Acyclic Graph (DAG) networks GoogLeNet, Inceptionv3, MobileNet-v2, Xception, AlexNet, VGG-16, VGG-19,  
DenseNet-201, Inception-ResNet-v2 and residual networks ResNet-18, ResNet-50 and ResNet-101 For the purpose of classification and comparison.

Maryam Bukhari \textit{\textit{et al.}} \cite{bukhari2022deep} by explicitly modeling channel interdependencies, she proposed squeeze and excitation learning, which iteratively executes recalibration on channel-wise feature outputs, highlighting the channel relationships on all levels of feature representation.

Ibrahim Abunadi \textit{et al.} \cite{abunadi2022multi} proposed three systems:  First, there is the artificial neural network, feed forward neural network, and support vector machine. These are based on hybrid characteristics that are derived using fuzzy color histogram, local binary pattern, and gray level co-occurrence matrix. The second system that is being suggested is made up of the transfer learning-based convolutional neural network models AlexNet, GoogleNet, and ResNet-18. Hybrid CNN-SVM technology is used in the third suggested solution.

Jyoti Rawat \textit{et al.} \cite{rawat2017classification} suggested methods comprise the following modules: segmentation (which divides each leukocyte cell into its nucleus and cytoplasm); feature extraction; feature dimensionality reduction (which maps the higher feature space to the lower feature space using principal component analysis); and classification (which uses standard classifiers such as k-nearest neighbor, probabilistic neural networks, adaptive neuro fuzzy inference systems, support vector machines, and smooth support vector machines).

Preetham Kumar \textit{et al.} \cite{suryani2017classification} presented a method for automatically identifying and dividing the white blood cell nucleus  from the images of the microscopic blood smear. K-Means is used for segmentation and clustering, and Support Vector Machine  with feature reduction is used for classification.

Esti Suryani \textit{et al.} \cite{kumar2017automatic} Utilizing the momentum backpropagation technique for feature extraction and Hue Saturation Intensity for Watershed Segmentation, automated AML detection was accomplished. The process of classification involves evaluating the numerical data input derived from the feature extraction outcomes that have been recorded in the database. 

T. T. P. Thanh\textit{\textit{et al.}} \cite{thanh2018leukemia} proposed a 
Convolutional Neural Network  based method in order to extract features from raw blood cell photos and perform classification in order to discriminate between normal and abnormal blood cell images.

Luis H.S. Vogado \textit{\textit{et al.}} \cite{vogado2018leukemia} extracted features from the photos using transfer learning, and then used three CNN architectures for further classification. The features were chosen based on their gain ratios and fed into the Support Vector Machine classifier.

Sarmad Shafique \textit{\textit{et al.}} \cite{shafique2018acute} deployed pretrained AlexNet which was fine-tuned. Data augmentation technique was applied to minimize overtraining, and performance over various color images was evaluated by comparing data sets with various color models.

Ibrahim Abdulrab Ahmed\textit{et al.} \cite{ahmed2023hybrid} suggested systems in which three CNN models (DenseNet121, ResNet50, and MobileNet) were used to extract WBC-only zones for additional analysis. The Principal Component Analysis approach selects highly representative features from the high features produced by CNN models, which are then supplied to the RF and XGBoost classifiers for classification.

\begin{table}[]
\caption{Summary of related works}
\resizebox{\textwidth}{!}{
\begin{tabular}{|p{2.5cm}|p{2.5cm}|p{3cm}|p{4cm}|p{1.5cm}|}
\hline
\textbf{Authors} & \textbf{Dataset} & \textbf{Segmentation Method} & \textbf{Classification model} & \textbf{Accuracy Rate} \\ \hline
\hline
Jyoti Rawat \textit{et al.} \cite{rawat2017classification} & ALL-IDB & \begin{tabular}[c]{@{}l@{}}Otsu’s Thresholding \end{tabular} & SVM & 89.00\% \\ \hline

Esti Suryani \textit{et al.} &Dr. Moewardi& Watershed Distance 
& \begin{tabular}[c]{@{}l@{}}Back Propagation\end{tabular} & \begin{tabular}[c]{@{}l@{}}94.285\%\\ \end{tabular} \\

\cite{kumar2017automatic} &  Hospital & 
Transform& &  \\\hline

Preetham Kumar \textit{et al.} \cite{suryani2017classification} & \begin{tabular}[c]{@{}l@{}}ASH\end{tabular} & k-means & SVM & 95\% \\ \hline

Maryam Bukhari \textit{et al.} \cite{bukhari2022deep} & ALL-IDB & Not Used & \begin{tabular}[c]{@{}l@{}}Squeeze and Excitation Learning\\CNN\end{tabular} & 98.3\% \\ \hline

Ibrahim Abunadi\textit{et al.} \cite{abunadi2022multi} & \begin{tabular}[c]{@{}l@{}}ALL-IDB\end{tabular} & Not Used & SVM & \begin{tabular}[c]{@{}l@{}}98.11\%\\ \end{tabular} \\ \hline

T. T. P. Thanh \textit{et al.} \cite{thanh2018leukemia} & \begin{tabular}[c]{@{}l@{}}ALL-IDB\end{tabular} & Not Used & CNN & 96.6\% \\ \hline

Sarmad Shafique \textit{et al.} \cite{shafique2018acute}  & ALL-IDB & Not Used & AlexNet & 96.06\% \\ \hline

Luis H.S. Vogado \textit{et al.} \cite{vogado2018leukemia} & Hybrid-Leukocyte database & Not Used & CNN-SVM & 99.00\% \\ \hline

 Ibrahim Abdulrab Ahmed\textit{et al.} \cite{ahmed2023hybrid} & C-NMC 2019 & Not Used & \begin{tabular}[c]{@{}l@{}}RF classifiers and XGBoost\end{tabular} & \begin{tabular}[c]{@{}l@{}} 99.1\%\end{tabular} \\ \hline

\end{tabular}
}
\label{tab:rw}
\end{table}




\section{Methodology}
\label{sec:Method}
\begin{figure}[H]
    \centering
    \includegraphics[scale=0.55]{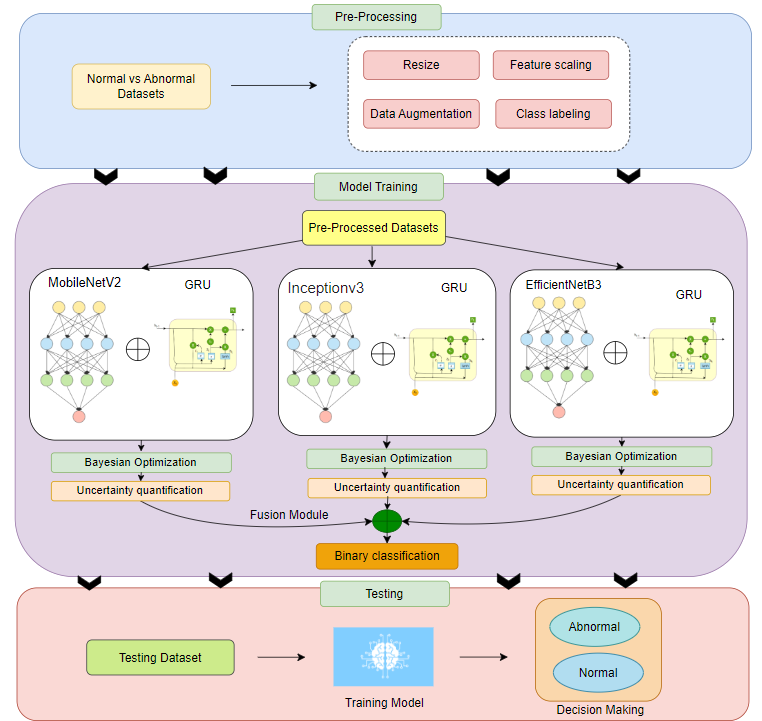} \hspace{0.5cm}
    \caption{ The basic block diagram of the proposed 
 model for ALL cancer detection.}
\label{fig:fusion}
\end{figure}

The proposed methodology's design is shown in Figure \ref{fig:fusion}.

\noindent
In summary, the proposed methodologies comprise the following steps. 
\begin{enumerate}
     \item The first step in the suggested framework is to take microscopic pictures of blood samples from two experimental datasets, namely \texttt{ ALL\_IDB1} and \texttt{ ALL\_IDB2}. For the preprocessing, since DL networks demand more data for training and better performance, the problem of inadequate data is resolved by employing data augmentation techniques followed by label resizing, class labeling and feature scaling.
     \item Next, Three distinct hybrid models were designed using different pretrained CNNs and Gated Recurrent Unit Network, such as MobileNetV2-GRU, InceptionV3-GRU, and EfficientNetb3-GRU. To avoid model overconfidence UQ approaches, such as Deep Ensemble was employed.
     \item Then, To optimize hyperparameters of these models Bayesian optimization technique was employed which has the benefits to reduce the search time and improves the model's performance by finding a better set of hyperparameters.
     \item Finally, feature fusion method based on sum rule at the score-leve is presented to identify leukemia based on input microscopic images of blood samples. 
\end{enumerate}

The subsequent methodological subsections provide a detailed explanation of each step:

\subsection{Image Dataset Collection}

The proposed study utilized images from the ALL-IDB, a publicly accessible collection that includes microscopic pictures
of blood samples. The dataset focuses on ALL, most lethal kind of leukemia disease. Lymphoma in each picture was identified and classified by lymphoma experts. All of the images were captured in JPG format with RGB color space (24 bits) and high resolution (2592 × 1944) with a Canon PowerShot G5 optical microscope. The ALL-IDB datasets come in two varieties: \texttt{ ALL\_IDB1} and \texttt{ ALL\_IDB2}.  There are 108 photos in the \texttt{ ALL\_IDB1} dataset: 59 of healthy individuals and 49 of lymphomas. Each image contains nearly 39,000 blood 
components classified by lymphoma experts. On the other hand, the \texttt{ ALL\_IDB2} dataset has 260 photos, 130 of which are of lymphomas and 130 of which are of normal cells. Regions from the \texttt{ ALL\_IDB1} dataset that have been cropped from blast cells and normal cells make up the \texttt{ ALL\_IDB2} dataset.



\subsection{Image Preprocessing}
The main goal of the image preprocessing is to
make the data suitable for deep learning models and improve the initial medical images by getting rid of noise, air
bubbles, and artifacts caused by gel that was sprayed before
the picture was taken. To attain a high classifying rate, we
removed artifacts and noise from the pictures. The elimination of this noise and artefacts was crucial to guaranteeing high-quality input
images since they could mask important features required for classification. The image was smoothed using
noise reduction algorithms, which also removed any random deviations that could confound the model.

\subsubsection{Data Augmentation}
\begin{figure}[H]
    \centering
\includegraphics[scale=0.5]{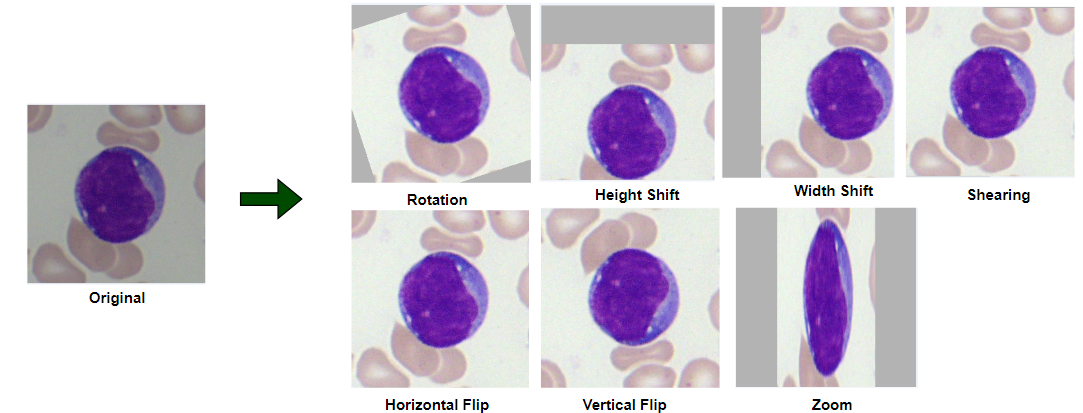} \hspace{0.5cm}
    \caption{Sample of leukemia images after each augmentation.}
    \label{fig:aug}
\end{figure}

CNN demonstrated state-of-the-art performance across a range of procedures. Nonetheless, the amount of training data has significant impact on CNN performance. Numerous techniques for data augmentation have reduced the network's error rate in image-based studies utilizing CNNs by offering hypothesis. Although there are many different microscopic blood sample images  utilized for this study, there are very few blood samples overall in both datasets. Thus, in order to overcome the problems of short dataset size and overfitting, we have employed a range of data augmentation approaches to artificially expand the quantity of data used for model training. Figure \ref{fig:aug} shows the sample augmented images that we generate by applying different
augmentation approaches. We used the following seven picture transformations to increase the sample number:
 \begin{enumerate}
   \item Rotation (45$^{\circ}$): To achieve this, the image was given a random rotation effect (left or
right). This procedure involves moving an image's pixel values left, right, up, and down in accordance with a degree value between 0 and 180. The degree value of 45$^{\circ}$ was chosen for this study in order to obtain various images. 
   \item Height Shift (20\%): The image pixels were arbitrarily moved up or down by 20\% to achieve the desired result.
   \item Width Shift (20\%): During this process, the pixel values were moved by 20\% to the right or left. After operations to change the height or width, a gap appeared in the opposite direction of
the shifted direction.
   \item Zoom (10\%): It brings objects in the picture closer in aspect. It was accomplished by enhancing the original image with new pixel values. The nearest value was found by examining the original pixel values before adding the values. 
    \item Horizontal Flip: In this operation, the pixels in the image were transferred horizontally from one half to the other half. The selection was made for the pixel values to travel arbitrarily.
    \item Vertical Flip: The image was split in half by a line drawn horizontally from its center.

    \item Shearing (20$^{\circ}$): To do this, the picture pixels were moved counter-clockwise in accordance with the degree of the set angle. This study's value was set at 20$^{\circ}$.
 \end{enumerate}

\subsubsection{Resizing}
In this study, resizing is another preprocessing technique. Since the proposed model requires equal image sizes to produce the intended training results, all of the photographs were resized to have the same dimensions (224 x 224).

\subsubsection{ Class Labeling}
We used label encoding techniques to transform the categorical data into numerical format because the labels in medical picture datasets are typically categorical (e.g., different forms of skin cancer). In order to enable the deep learning model to process and learn from the labels during training, each distinct category was given a numerical value. The mapping between numerical and categorical data for the class labeling is as follows:  the label 0 denotes people without ALL, and the label 1 denotes individuals who have ALL.

\subsubsection{Feature scaling}
The feature scaling method is well
recognized in the field of machine learning and pattern recognition for being used to normalize data. To avoid outliers,
this approach sets all data items to the same scale and thus
improves prediction quality. As the features in cancer datasets
have a high variance, one of the pre-processing strategies
for normalization is feature scaling. We divided the grayscale
value of an image by 255 to standardize our image pixels
between 0 and 1. As a result, the numbers will be relatively
little, and the computation will be simpler and faster.



\subsection{Pretrained Networks}
There are several pretrained CNNs that have been made accessible to the public that were trained on over a million pictures. Once these pretrained CNNs have been modified and optimized for a standard classification task, they can be trained using the new dataset. The proposed work 
used the pretrained networks MobileNetV2, InceptionV3 and EfficientNetB3.

 With its inverted residual and linear bottleneck layer, MobileNetv2's CNN layer delivers improved accuracy and performance for embedded and mobile video applications. In general,  MobileNetV2's architecture consists of a fully convolution layer with 32 filters at the beginning, followed by 19 residual bottleneck layers. This architecture lowers the network's complexity cost and is specifically designed for devices with little processing capability. Its capacity to attain competitive accuracy in comparison to larger and more computationally expensive models is another impressive feature of this model. Finally, the modest size of the model allows for faster inference times, which makes it appropriate for real-time applications.

 The InceptionV3 is a DL model based on CNN with 42 layers, which is used for image classification. It is developed by Google, has a high image classification performance. It utilized multiple approaches  for optimizing the network in order to improve model adaptation. While maintaining the same speed, it is more efficient and has a deeper network. It is computationally less expensive improving the decision functions and allowing the network to converge quickly.  

 The EfficientNet family is based on a new way for scaling up CNN models. It makes use of a basic compound coefficient that is quite effective. This architecture is founded on the premise that scaling up the process for width, height, and depth for neural networks must be balanced.  Effectively optimizing every aspect of the network in relation to the resources at hand enhances overall performance. We suggest utilizing the EfficientNetB3 due to its ability to offer a favorable balance of processing power, precision, and execution duration. The design consists of 26 convolution blocks nested after a convolution layer with swish activation.



\subsection{Gated Recurrent Unit(GRU) Network}
\begin{figure}[H]
    \centering
    \includegraphics[scale=0.5]{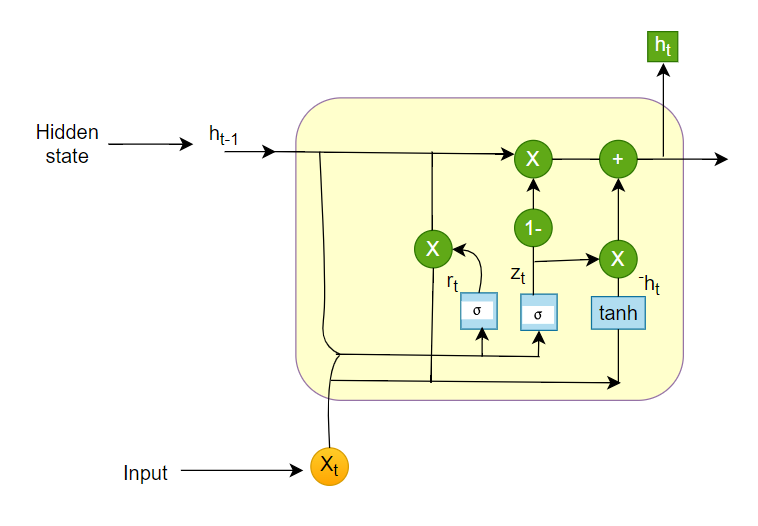} \hspace{0.5cm}
    \caption{The basic structure of the GRU model.}
    \label{fig:gru}
\end{figure}

The goal of GRU is to resolve the vanishing gradient issue that arises when using a conventional recurrent neural network, as shown in Figure \ref{fig:gru}. In 2014, Kyunghyun Cho at el. presented it\cite{khdhir2023pancreatic}. 
Due to its internal cell state and three major gates, GRU is more efficient than LSTM. The data is kept within the GRU in a secret state. GRU makes use of update and reset gates to solve the vanishing gradient problem of typical recurrent neural network. These two vectors essentially decide which data should be sent to the output. What sets them apart is their capacity to be trained to remember past data without deleting it or removing information that is irrelevant to the prediction. The update gate (U) provides both forward and backward information, while the reset gate (R) presents prior knowledge. The reset gate is utilized by the current memory gate to retain the necessary data from the computer's prior state. It's feasible to give the input nonlinearity and zero-mean features at the same time by using an input modulation gate.
The statement that follows states that the basic GRU of reset and updated gates can be mathematically described as follows:
\[ \( R_{\text{t}}\) = \sigma (\( X_{\text{t}}\).\( w_{\text{xr}}\) + \( H_{\text{t-1}}\).\(w_{\text{hr}}\) + \( b_{\text{r}}\)) \]
\begin{equation*}            
\( U_{\text{t}}\) = {\sigma (\( X_{\text{t}}\).\( w_{\text{xz}}\) +\( H_{\text{t-1}}\).\(w_{\text{hz}}\) + \(b_{\text{z}}\))}
\end{equation*}
\noindent
where \( w_{\text{rx}}\) and \( w_{\text{xz}}\) are weight parameters and \( b_{\text{r}}\) and \( b_{\text{z}}\)
represent bias vector\cite{ahmad2022novel}.

\subsection{Hybrid Models}
\begin{figure}[!htbp]
    \centering
    \subfigure[MobileNetV2-GRU]{\includegraphics[scale=.25]{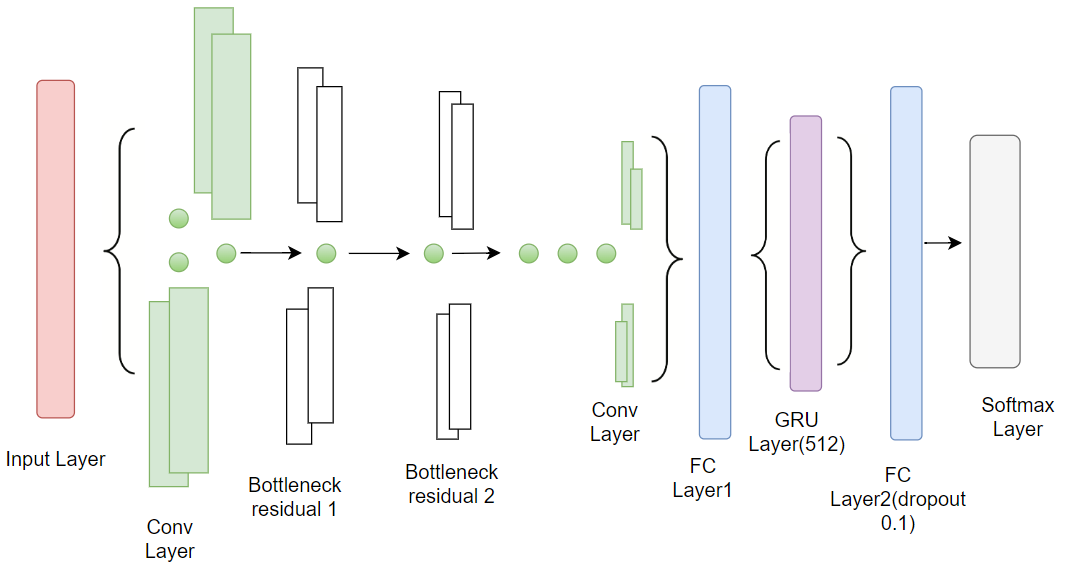}}\label{igru}\hspace{0.5cm}
    \subfigure[InceptionV3-GRU]{\includegraphics[scale=.25]{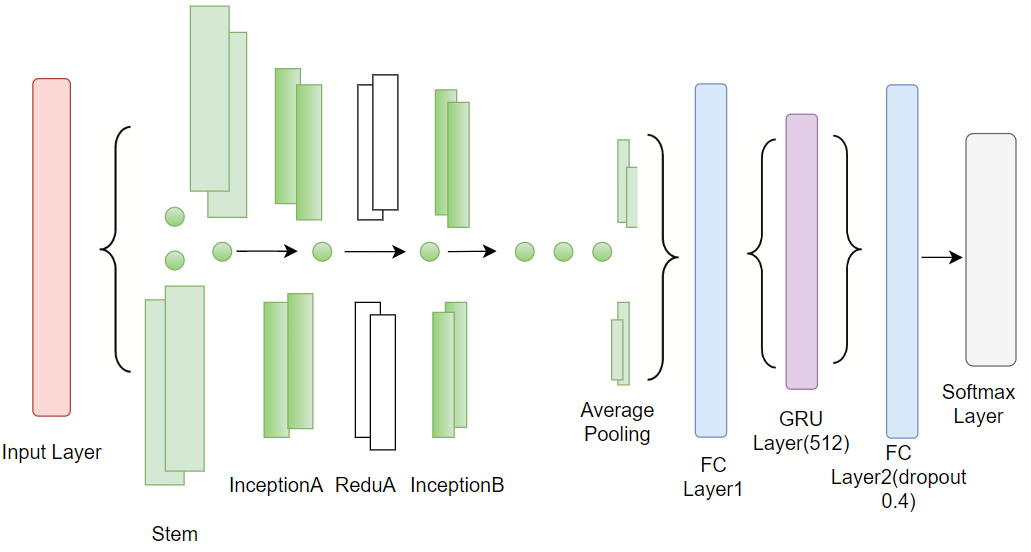}}
    \subfigure[ EfficientNetB3-GRU]{\includegraphics[scale=.25]{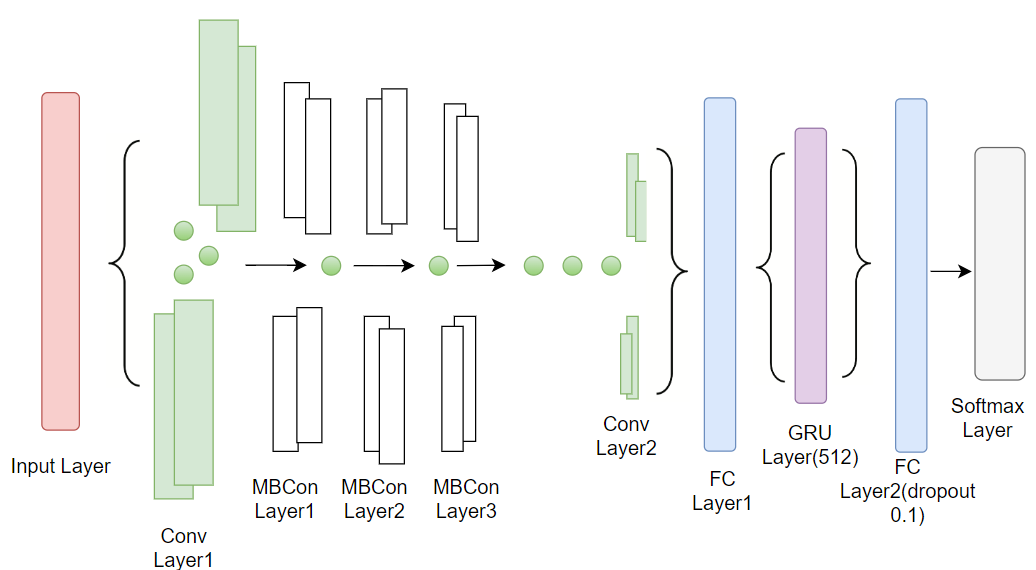}}
    \caption{The general structure of the proposed Hybrid models.}
    \label{fig:hybrid}
\end{figure}

In this study, we propose the InceptionV3-GRU, EfficientNetB3-GRU, and MobileNetV2-GRU models for ALL classification. Figure \ref{fig:hybrid}(a) depicts the structure of the suggested MobileNetV2-GRU model. When the image was first collected, it had the dimensions (224, 224, 3) with 224 pixels for height and 224 pixels for width in RGB, and three channels. The convolutional layers of the suggested model process the image  in order to extract its features. In addition, the convolutional layer's stride and kernel size were 1 and (3 × 3), respectively. The training parameter was significantly reduced after each conventional and max-pooling layer, and was followed by dropout and the activation function (ReLU). The data must be assembled into an array following the training of the max-pooling and traditional layers in order to be used as input for the  fully connected layer that is built by flattening with training parameters of size (5,5) and features map of value 1536. Following the completion of all convolutional layers, 1,024 feature maps were produced using the dropout method. To tackle the vanishing gradient problem, a GRU model with a fully connected layer made up of 512 neurons was used. Two fully connected layers were also employed after the GRU model. Using the last linked layer, softmax operations were finally implemented.

Figure \ref{fig:hybrid}(b) displays the structure and parameters used in the InceptionV3-GRU model. At first, the image's input shape was (224, 224, 3) size, with 224 pixels for the image's height and 224 pixels for its width in RGB, and three channels. The input shape goes through the convolutional layers of the suggested model in order to extract the features. Furthermore, the convolutional layer's stride and kernel size were (3 × 3) and 1. The training parameter was significantly reduced after each conventional and max-pooling layer, and was followed by dropout and the activation function (ReLU). The data must be assembled into an ID array following the training of the max-pooling and traditional layers in order to be used as input for a fully connected layer that is built by flattening with training parameters of size (7, 7)  and features map of value 512. Following the completion of all convolutional layers, 1,024 feature maps were produced using the dropout method. To tackle the vanishing gradient problem, a GRU model with a fully connected layer made up of 1024 neurons was used. Two fully connected layers were also employed after the GRU model. Using the last linked layer, softmax operations were finally implemented.

\begin{figure}[t]
    \centering
    \includegraphics[scale=0.40]{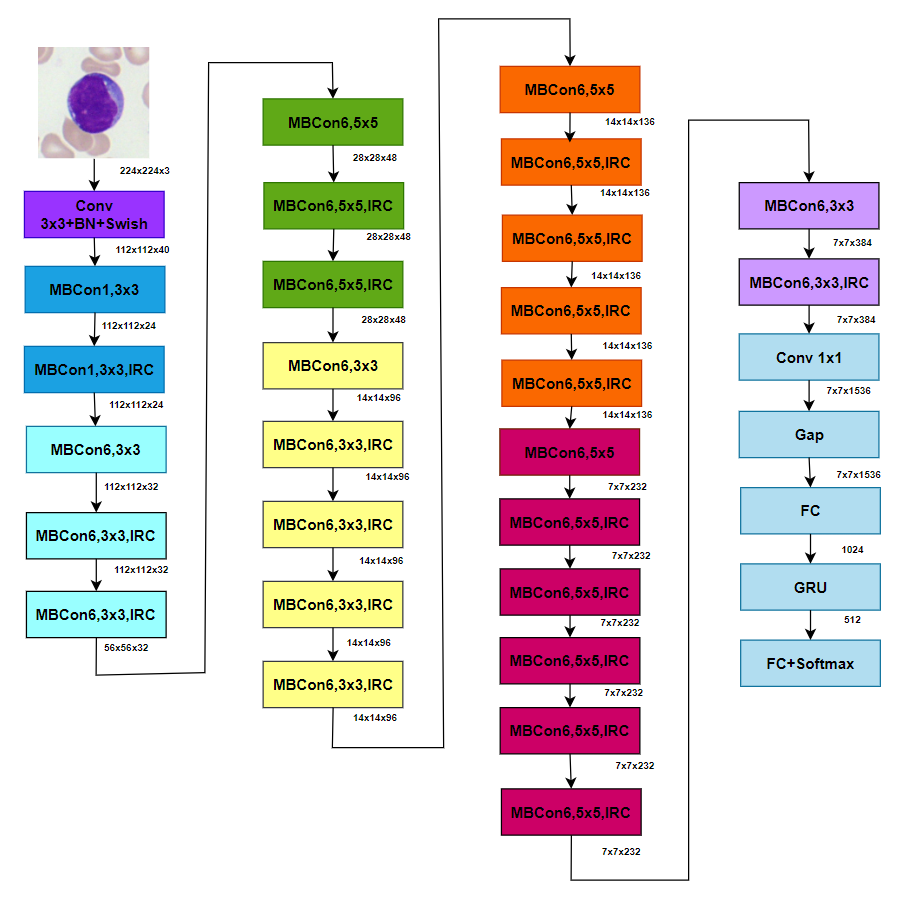} \hspace{0.5cm}
    \caption{EfficientNetB3-GRU network architecture.}
    \label{fig:eb}
\end{figure}

The structure and parameters used in the EfficientNetB3-GRU model are displayed in Figure \ref{fig:hybrid}(c). Initially, the image's input shape was (224, 224, 3) size, with 224 pixels for the image's height and 224 pixels for its width in RGB, and three channels. The input shape goes through the convolutional layers of the sug-
gested model in order to extract the features. Furthermore, the convolutional
layer’s stride and kernel size were (3 × 3) and 1. Activation function (ReLU) and drop out were implemented after each conventional and max-pooling layer, with a significant decrease in the training parameter.The data must be assembled into an ID array following the training of the max-pooling and traditional layers in order to serve as input for a fully connected layer that is created by flattening with training parameters of size (7, 7)  and features map of value 1536. Once all of the convolutional layers had been processed, 1,024 feature maps were produced using the dropout method. To overcome the vanishing gradient problem, a fully connected layer of 512 neurons in a GRU model was used. There were two fully connected layers utilized after the GRU model. Lastly, the last linked layer was used to implement softmax operations. To more clearly express the internal composition of
its network, the structure of each part of the EfficientNetB3-GRU model is
shown in Figure \ref{fig:eb}.

\subsection{Hyperparameter Tuning}

Bayesian Optimization (BO) is a probabilistic method utilizes for the hyperparameter tuning. The Bayesian theorem serves as its foundation. The two main parts of BO are the acquisition function, which is optimized for selecting the next sample location, and the surrogate model, which is a prior distribution that models the unknown objective function. The surrogate model generates a posterior probability distribution about probable values for H($\theta$) at a candidate configuration $\theta$. The main concept is to update this posterior distribution every time we observe H at a new point $\theta$. It develops a probabilistic surrogate model, frequently by taking into account a tree-based model or a Gaussian process. It applies a Gaussian process prior over the optimization functions. The Gaussian Process is fully characterized by-
\begin{itemize}
    \item A mean function $\mu$($\theta$) : $\Theta$ $\rightarrow$ R to get assumptions about the optimized function.
    \item  A definite positive covariance function, also called kernel $k$($\theta$,$\theta'$) : $\Theta$$^2$ $\rightarrow$ R.
\end{itemize}
The following mathematical expression states that the function $H(\theta)$ follows a Gaussian process with mean function $\mu(\theta) $
and covariance (kernel) function $k(\theta$,$\theta')$.

\begin{equation*}
     H(\theta) \sim GP(\mu(\theta)); k(\theta,\theta'))
 \end{equation*}
The first set of $k$ configurations is used by the BO algorithm $(\theta_i)^k_{i=1}$
and their associated function values $(y_i)^k_{i=1}$
with $y_i = H(\Theta_i)$. The GP model is updated using the Bayes rule at each iteration, $t \in \{ k+1,....,N \}$, to produce the posterior distribution conditioned on the current training set, $S_t = \{(\theta,y_i)\}^t_{i=1}$
which contains the previously evaluated configurations and observations.
If a new point $\theta_{t+1}$ 
is selected and evaluated to provide an observation $y_{t+1} = H (\theta_{t+1})$
, we add the new pair $(\theta_{t+1}, y_{t+1})$
to the current training set $S_t$ , obtaining a new training set for the next iteration $S_{t+1} = S_t\cup(\theta_{t+1}, y_{t+1})$ \cite{galuzzi2020hyperparameter}. 
In order to choose the next candidate point to be evaluated, an auxiliary optimization problem of the following general form should be solved:
\begin{equation*}
     \theta_{t+1} = arg max U_t(\theta;S_t),\theta \in \Theta
 \end{equation*}
where $U_t$ is the acquisition function that needs to be maximized.The BO process continues to traverse until the maximum value has been reached. By utilising all the data it receives from the optimization history, BO makes this search effective\cite{shahriari2015taking}\cite{canayaz2022covid}. The BO pseudocode is provided in Algorithm \ref{alg:alg1}.


\begin{algorithm}
\caption{Bayesian optimization}
\label{alg:alg1}

\begin{algorithmic}[1] 
\State Initialize data $y_0$ using initial design
\For n = 1,....,$n_{max}$ do
   \State Find $\theta_n$ $\in$ $\Theta$ by optimizing the acquisition function $U_n$,
         
         $\theta_{t+1}$ = arg max $U_t$($\theta$;$S_t$)
   \State Evaluate the objective function $y_n$ = H($\theta_n$)
   \State  Augment the data $S_n$ = $S_{n-1}$ $\cup$ \{$\theta_n$, $y_n$\}
   \State Update the surrogate model
\EndFor

\end{algorithmic}
\end{algorithm}


\subsection{Uncertainty Quantification}
\begin{figure}[t]
    \centering
    \includegraphics[scale=0.55]{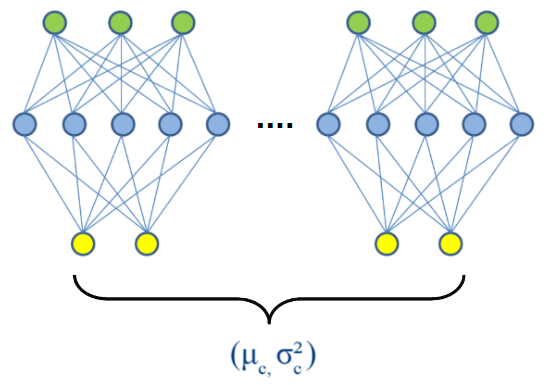} \hspace{0.5cm}
    \caption{Deep ensembles uncertainty quantification.}
    \label{fig:DE}
\end{figure}

Uncertainty Quantification (UQ) is  essential  for an accurate application of ML and DL methods. The confidence of the outcomes produced by these techniques can be raised using a UQ estimation. Deep Ensembles(DE) were proposed by Lakshminarayanan \cite{lakshminarayanan2017simple} , which is easy to implement, easily parallelizable, require minimal hyperparameter adjustment, and produce high-quality prediction uncertainty estimation.
 The technique is used to train a group of networks using various random initializations. Every network operates in a similar way when there is enough training data available, but when there is none at all, the outcomes are quite different. Instead of training a single network, we will train an ensemble of \(M\) networks, using various random initializations as shown in figure \ref{fig:DE} . For a final prediction, we now take all networks and aggregate their output into a Gaussian mixture distribution, from which we can derive the variance estimation $\sigma_c^2$ and single mean $\mu_c$ which is mathematically expressed as follows.

\begin{equation*}
    \hat\mu_c(x) = \frac{1}{M} \sum_{i=1}^{M} \hat\mu_i(x)
    \end{equation*}
\begin{equation*}
    \\\hat \sigma_c^2(x) = \frac{1}{M} \sum_{i=1}^{M} \hat\sigma^2_i(x) + \left[ \frac{1}{M} \sum_{i=1}^{M} \hat\mu^2_i(x)-\hat\mu_c^2(x) \right]
\end{equation*}

\noindent
We evaluate DE on the ALL-IDB1, ALL-IDB2 and combine datasets  using the InceptionV3-GRU, EfficientNetB3-GRU, and
MobileNetV2-GRU networks. An additional advantage of employing an ensemble is that it makes it simple to pinpoint training samples where the various networks dispute or agree the most. 
 This disagreement presents an additional helpful qualitative method of evaluating prediction uncertainty.




\subsubsection{Feature Fusion Hybrid Model}
Feature fusion methods are computer programs that use scores from different models to make a decision.  A single scalar score is often generated as a result of this consolidation process. It typically utilizes relatively easy fusion operators and doesn't require a lot of computing. It has been discovered that the best outcomes are obtained when combining various score-level fusion techniques employing outputs with comparable performances. A reasonably simple score-level fusion technique that works directly with raw score data is the sum rule. The  sum rule is shown as follows:
\begin{equation*}
     f_s = {x_i + x_m+ x_e}
 \end{equation*}

According to the sensitivity analysis, the sum rule is the most durable method for error estimation. It has been found that the sum rule performs better than most other state-of-the-art score-level fusion algorithms.

The final decision of the proposed  method is made by combining the outputs generated from three separate DL models : MobileNetV2-GRU($x_m$), EfficientNetB3-GRU($x_e$) and InceptionV3-GRU($x_i$) models for the accurate classification of ALL cancer as shown in Figure \ref{fig:fusion}. We separately train  
each hybrid models using input images. During training, we use the BO to determine the optimal of hyperparameters. The BO application optimizes achieving better results at the test stage. Well-known uncertainty method, i.e., DE is employed and tested on ALL-IDB datasets to reduce models uncertainty. 
 
  For each input image, three anticipated
probability ratings are generated, and the input image is
classified into either the ALL or noncancer class based on
the highest probability score.
To combine the results from the three models, we used
the  sum rule, taking into account the parallel architecture in the suggested system. This approach provides
hematologists with a high level of confidence to make the final
judgment and accurately discriminate between cancer-infected
and healthy patients. 
Our proposed hybrid framework significantly improves the
accuracy of ALL cancer classification when compared to  MobileNetV2, EfficientNetB3 and InceptionV3 models. Our intention is not only to enhance 
the overall prediction performance, also at defining a model with higher confidence in its predictions. Clearly, the suggested model can handle  uncertainties in its predictions. This is essential because it makes it easier to identify leukemia cancer patients from healthy ones.  With our
feature fusion approach, we aim to provide a powerful tool
for hematologists to improve their diagnoses and ultimately
improve patient outcomes. In our proposed hybrid model, several of the parameters are
amended as follows Table \ref{tab:hyp}.

\begin{table}[!htbp]
\caption{Hyperparameters of different deep learning architectures after using Bayesian Optimization.}
\centering
\begin{tabular}{p{2.5cm}p{2.5cm}p{2.5cm}p{2.5cm}} 
\hline
\multicolumn{1}{l}{Hyperparameter} & \multicolumn{1}{l}{InceptionV3}   & \multicolumn{1}{l}{MobileNetV2}      & \multicolumn{1}{l}{EfficientNetB3} \\ \hline
\multirow{5}{*}{}Units   & 512      & 512   & 256      \\ 
     Activation & Softmax   & Softmax & Softmax     \\
    Optimizer & SGD & SGD & RMSprop   \\ 
     Learning rate & 0.01 & 0.01 & 0.0001   \\ 
        Momentum & 0.3 & 0.9 & 0.5   \\ 
        Dropout rate & 0.4 & 0.1 & 0.1 \\
      \hline
 
\end{tabular}

\label{tab:hyp}    
\end{table}

\section{Experimental Results Analysis}

\subsection{Environment Setup}
This section provides the suggested model's results in a range of experimental contexts, along with comparisons and discussions. Furthermore, the recommended model is implemented using Python in combination with the Keras deep learning framework, and the simulations are carried out on a powerful computer system equipped with an Intel Core i7 processor, 16 GB of RAM, and an NVIDIA GeForce MX330 GPU. Two datasets are used in the performance and setup of the experiment. The BO technique defines every parameter, and the optimal combination of parameter values is used to report the findings.
\subsection{Evaluation Criteria}
In order to evaluate the performance of our proposed model, we employ various performance metrics such as accuracy, precision, recall, F1-measure, specificity, and a confusion matrix. The confusion matrix is made up of four main terms: True Negative (TN), True Positive (TP), False Negative (FN), and False Positive (FP), which offer valuable information on the overall effectiveness of the model. The performance metrics utilized in this study are listed below:

\textbf{Accuracy : }
This metric represents the total number of classes—that is, Acute Lymphoblastic Leukemia (ALL) and not Acute Lymphoblastic Leukemia (not ALL)—that the trained model was able to predict correctly out of all classes. This measure depicts the proportion of patients with leukemia and those without. The higher the value of
accuracy, the more accurate is the model. The
equation of accuracy is shown in the following formula:
\begin{equation*}
    Accuracy= \frac{TP + TN}{TP + TN + FP + FN}
\end{equation*}

\textbf{Precision: }
This metric determines the percentage of true positives among all positive cases. What matters in the case of leukemia disease is the model's accuracy in identifying those people who are afflicted.
Mathematically, it is defined as in the following equation:
\begin{equation*}
     Precision = \frac{TP}{TP+FP}
\end{equation*}

\textbf{Recall : }
The recall evaluates how well, given the total relevant data, the model highlights the patients with leukemia. The following formula is used to calculate it:
\begin{equation*}
     Recall = \frac{TP}{FN + TP}
 \end{equation*}

 
 \textbf{F1 measure : }
This metric integrates the recall and precision numbers to determine the model's overall efficiency.
 \begin{equation*}
    F_1 = 2 \times \frac{Precision \times Recall}{Precision + Recall}
\end{equation*} 

\textbf{Specificity: } Specificity is the metric that evaluates a model’s ability to predict true negatives of each available category. These metrics apply to any categorical model. The equations for calculating this metrics are as follows-

\begin{equation*}
      Specificity =\frac{TN}{FP + TN}
 \end{equation*}

\textbf{Confusion Matrix: } The confusion matrix is a technique for assessing performance in the form of a table that incorporates information about both actual and expected classes. The dimension of the confusion matrix would be nxn if the proposed problem to be examined has an n row, with the rows representing the actual row and the columns representing the predicted row. For two or more classes, the matrix depicts actual and anticipated values.

\textbf{Reciever Operating Characteristics: }One useful method for predicting the possibility of a binary result is the ROC curve. On a probability curve, the true-positive rate (TPR) vs the false-positive rate (FPR) at various thresholds are displayed. The ROC curve illustrates the trade-off between specificity and sensitivity. The true positive indicates how effectively the model predicts the positive class when the actual result is positive. The shape of the curve provides a variety of information, such as the things that we would find most important for a certain situation, the expected false positive rate and false-negative rate.
\subsection{Results}
This section presents the classification results obtained with the 
proposed feature fusion hybrid model for the datasets \texttt{ ALL\_IDB1} , \texttt{ ALL\_IDB2} and \texttt{ ALL\_IDB1}+\texttt{ ALL\_IDB2} separately. To ensure the accuracy and generalizability of our model,
we employ a rigorous methodology that involves an 80\%
training and 20\% testing dataset split. Moreover, to prevent
overfitting, we randomly select 20\% of the training set as validation set, which is used to determine the optimal weights
that produce the lowest error.

\subsubsection{Results of the \texttt{ ALL\_IDB1} Database}
\begin{figure}[H]
    \centering
    \subfigure[ALL-IDB1]{\includegraphics[scale=.5]{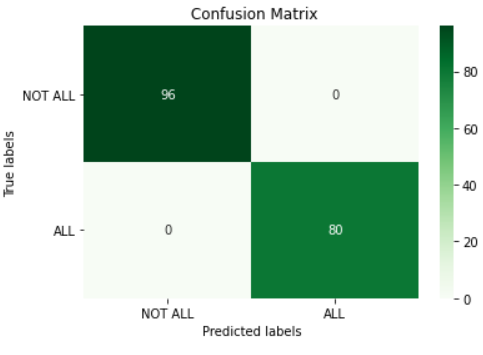}}\label{igru}\hspace{0.5cm}
    \subfigure[ALL-IDB2]{\includegraphics[scale=.5]{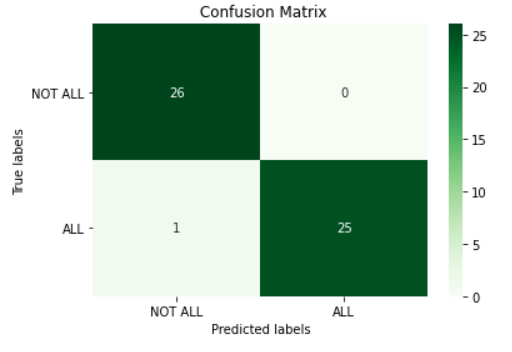}}
    \subfigure[ ALL-IDB1 + ALL-IDB2]{\includegraphics[scale=.5]{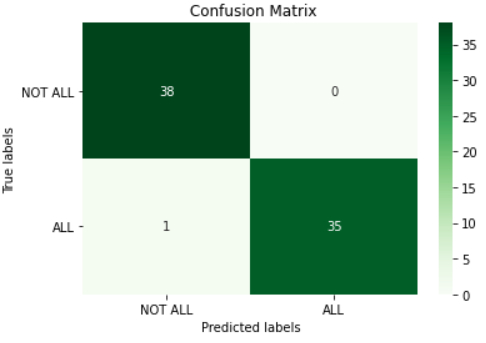}}
    \caption{Confusion Matrix of the proposed model for ALL-IDB1, ALL-IDB2 and ALL-IDB1 + ALL-IDB2 datasets.}
    \label{fig:con}
\end{figure}

The suggested model is initially validated using the \texttt{ ALL\_IDB1} database in order to evaluate its performance. As mentioned earlier, there are extremely few microscopic blood samples available for training that are both ALL and not ALL. Therefore, in order to enhance the total amount of blood samples for training, we have used data augmentation approaches. As a result, the train set has an acceptable amount of samples, which are used to train the model after data augmentation. Table \ref{tab:all1} provides the total number of blood samples used for training and testing for both ALL and non-ALL classes. The model is trained using these enhanced photos.
The recommended model extracts features of leukemic cells from the max-pool and convolution layers. GRU is applied to the reduced gradient vanishing problem, and UQ is applied to each hybrid model to boost model confidence. As illustrated in table \ref{tab:hyb}, it is observed that EfficientNetB3+GRU+DE  method gives an accuracy of 95.45\%,
 MobileNetV2+GRU+DE method gives an accuracy of 100\% and InceptionV3+GRU+DE gives 100\% accuracy. The parallel
architecture was taken into account in the proposed system,
giving hematologists a high level of confidence to make the
final decision that
correctly diagnosing leukemia patients with 100\% accuracy shown in Table \ref{tab:Dataset}. Moreover, the F1Score values, precision, and recall of the extra evaluation metrics are 100\%,
100\% and 100\%, in that order. Moreover,  we have also plotted the confusion matrix of the experiment, as illustrated in Figure \ref{fig:con} (a). The confusion matrix shows the overall model efficiency for every class category in the dataset. The results clearly demonstrate that the suggested model performs better when it comes to dividing the blood samples into ALL and not ALL classes.
\begin{table}[!htbp]
\caption{Training samples and testing samples on the ALL-IDB1 dataset.}
\centering
\begin{tabular}{p{2.5cm}p{2.5cm}p{2.5cm}p{2.5cm}} 
\hline
\multicolumn{1}{l}{\textbf{Sample Type}} & \multicolumn{1}{l}{\textbf{NOT ALL}}   & \multicolumn{1}{l}{\textbf{ALL}}      & \multicolumn{1}{l}{\textbf{Total}} \\ \hline
\multirow{5}{*}{}\textbf{Training}   & 376      & 313   & 689      \\ 
     \textbf{Test} & 12   & 10 & 22     \\
    \textbf{Total}& 388 & 323 &711   \\ 
 
      \hline
 
\end{tabular}

\label{tab:all1}    
\end{table}

\begin{table}[ht]
\caption{Obtained results of different deep learning and  UQ method for the ALL-IDB1, ALL-IDB2 and ALL-IDB1 + ALL-IDB2 datasets.}
\centering

\begin{tabular}{p{3cm}p{5cm}p{2cm}} 
\hline
\multicolumn{1}{l}{Dataset} & \multicolumn{1}{l}{Method}      & \multicolumn{1}{l}{Accuracy} \\ \hline

\multirow{5}{*}{ALL-IDB1 }        & EfficientNetB3+GRU+DE   & 95.45\%       \\ 
       & MobileNetV2+GRU+DE & 100\%     \\ 
           
     & InceptionV3+GRU+DE    & 100\%       \\ 
     
    \hline
          
\multirow{5}{*}{ALL-IDB2 }         & EfficientNetB3+GRU+DE & 92.31\%     \\ 
      & MobileNetV2+GRU+DE & 94.23\%    \\ 
    & InceptionV3+GRU+DE & 92.31\%    \\

    \hline
    
\multirow{5}{*}{ALL-IDB1 +}        & EfficientNetB3    & 97.30\%    \\ 
\multirow{5}{*}{ALL-IDB2}        & +GRU+DE     &    \\ 

 & MobileNetV2+GRU+DE  & 95.95\%       \\
 & InceptionV3+GRU+DE  & 94.59\%       \\
            \hline  
\end{tabular}

\label{tab:hyb}    
\end{table}

\begin{table}[htbp]
\caption{Results of the proposed model on both ALL1-IDB1, ALL-IDB2 and ALL-IDB1 + ALL-IDB2 datasets}
\label{tab:Dataset}
\resizebox{\textwidth}{!}{
\centering
\begin{tabular}{|l|l|l|l|l|}
\hline
\textbf{Dataset} & \textbf{Accuracy(\%)} & \textbf{Precision(\%) } & \textbf{Recall(\%)} & \textbf{F1-score(\%)} \\ \hline
\multirow{2}{*}{} 
 {ALL-IDB1}& 100 & 100 & 100 & 100 \\ \hline
\multirow{11}{*}{}
{ALL-IDB2}& 98.08 & 100 & 96.15 & 98.04 \\ \hline
\multirow{11}{*}{}
{ALL-IDB1 + ALL-IDB2}& 98.64 & 100 &  97.22& 98.59 \\ \hline
\end{tabular}
}
\end{table}



\subsection{Results of the \texttt{ ALL\_IDB2} Database}
In the second validation stage, we utilized the second dataset, \texttt{ ALL\_IDB2}. The total number of microscopic blood samples in this dataset is likewise very inadequate for training the suggested model. Therefore, this dataset undergoes the same data augmentation process. Table \ref{tab:all2} presents the total count of blood samples used for training and testing, both for ALL and not ALL classes.  Afterwards, the model is trained using the train set that has an excessive amount of apparent variety. 
Additionally, the suggested model's hyperparameters are all the same as the ones we initially defined with the first database. 
Table \ref{tab:Dataset} displays how the suggested model performed on this database. The convolutional and pooling layers yield the features of microscopic images of blood samples whose learning increases by using GRU to solve the reduced gradient vanishing problem, and each hybrid model is given a UQ application to increase model confidence. As shown in Table \ref{tab:hyb}, it is observed that EfficientNetB3+GRU+DE  method gives an accuracy of 92.31\%,
 MobileNetV2+GRU+DE method gives an accuracy of 94.23\% and InceptionV3+GRU+DE gives 92.31\% accuracy. When compared to EfficientNetB3+GRU+DE,  MobileNetV2+GRU+DE or InceptionV3+GRU+DE model alone, the accuracy
of the suggested hybrid system has been significantly improved. Table \ref{tab:Dataset} demonstrates that the model is performing well on this dataset as well. The model's  yielded an overall accuracy of almost 98.08\%, while its precision, recall, and F-Score values were 100\%, 96.15\% and 98.04\% respectively. It's also important to note that, although we used the complete microscopic images of the blood samples in the first experiment, the cell areas in this dataset have been cropped. The optimal outcomes on both kinds of image settings are displayed by the suggested model. The confusion matrix for the experiment using cropped cell pictures is then also created. Figure \ref{fig:con}(b) displays the experiment's confusion matrix.
\begin{table}[!htbp]
\caption{Training samples and testing samples on the ALL-IDB2 dataset.}
\centering
\begin{tabular}{p{2.5cm}p{2.5cm}p{2.5cm}p{2.5cm}} 
\hline
\multicolumn{1}{l}{\textbf{Sample Type}} & \multicolumn{1}{l}{\textbf{NOT ALL}}   & \multicolumn{1}{l}{\textbf{ALL}}      & \multicolumn{1}{l}{\textbf{Total}} \\ \hline
\multirow{5}{*}{}\textbf{Training}   & 832     & 832   & 1664      \\ 
     \textbf{Test} & 26   & 26 & 52     \\
    \textbf{Total} & 858 & 858 
 &1716   \\ 
    &       \\ 
      \hline
 
\end{tabular}

\label{tab:all2}    
\end{table}



\subsection{Results of Combining Both Datasets}
In the
third experiment,To enhance diversity and boost the quantity of test photos, we have included the microscopic images from both datasets. Similarly, we also used data augmentation in this experiment to raise the training size. Table \ref{tab:comb} displays the total number of training and testing samples for ALL and not ALL classes.
The train set with significant data augmentation is provided as an input to the suggested model, as was previously done for each of the datasets separately. The leukemic and normal cell features are extracted by the model from cropped and full-size images, respectively. It's also encouraging how accurate this scenario turned out to be that can be perceived from Table \ref{tab:hyb} and Table \ref{tab:Dataset}. Likewise, there has been an improvement in recall, precision, and F-Score values. In the third experiment, recall is accomplished on ALL classes at 97.22\%, but in the second experiment, recall is only 96.15\%.
In addition, as illustrated in Figure \ref{fig:con}(c) , the experiment's confusion matrix is also plotted. 
\begin{table}[!htbp]
\caption{Training samples and testing samples by combining both datasets.}
\centering
\begin{tabular}{p{2.5cm}p{2.5cm}p{2.5cm}p{2.5cm}} 
\hline
\multicolumn{1}{l}{\textbf{Sample Type}} & \multicolumn{1}{l}{\textbf{NOT ALL}}   & \multicolumn{1}{l}{\textbf{ALL}}      & \multicolumn{1}{l}{\textbf{Total}} \\ \hline
\multirow{5}{*}{}\textbf{Training}   & 1208     & 1145   & 2353     \\ 
     \textbf{Test} & 38   & 36 & 74     \\
    \textbf{Total} & 1246 & 1181 
 &2427   \\ 
    &       \\ 
      \hline
 
\end{tabular}

\label{tab:comb}    
\end{table}

\begin{figure}[H]
    \centering
    \subfigure[ALL-IDB1]{\includegraphics[scale=.5]{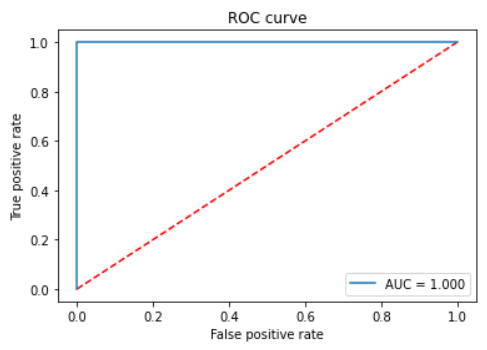}}\label{igru}\hspace{0.5cm}
    \subfigure[ALL-IDB2]{\includegraphics[scale=.5]{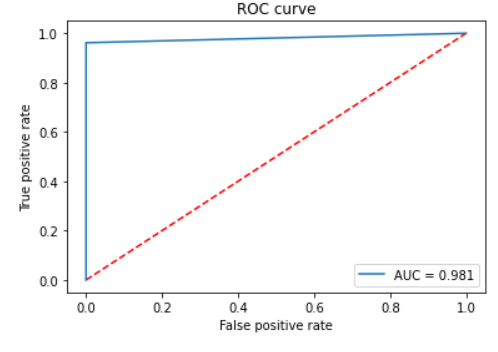}}
    \subfigure[ ALL-IDB1 + ALL-IDB2]{\includegraphics[scale=.5]{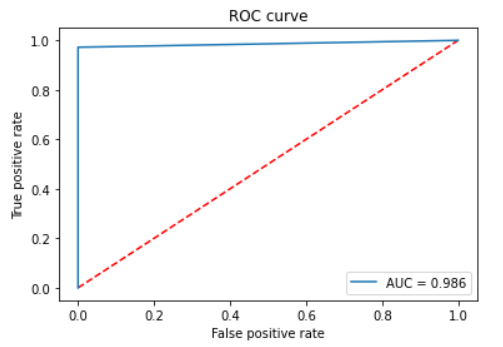}}
    \caption{ROC curve of the proposed model for ALL-IDB1, ALL-IDB2 and ALL-IDB1 + ALL-IDB2 datasets.}
    \label{fig:roc}
\end{figure}

Furthermore, receiver operating curves (ROC) are also plotted for each database. The ROC curves
are depicted in Figure \ref{fig:roc}(a), \ref{fig:roc}(b) and \ref{fig:roc}(c) for ALL-IDB1, ALL-IDB2 and combine datasets respectively. 

\begin{figure}[H]
    \centering
    \includegraphics[scale=0.3]{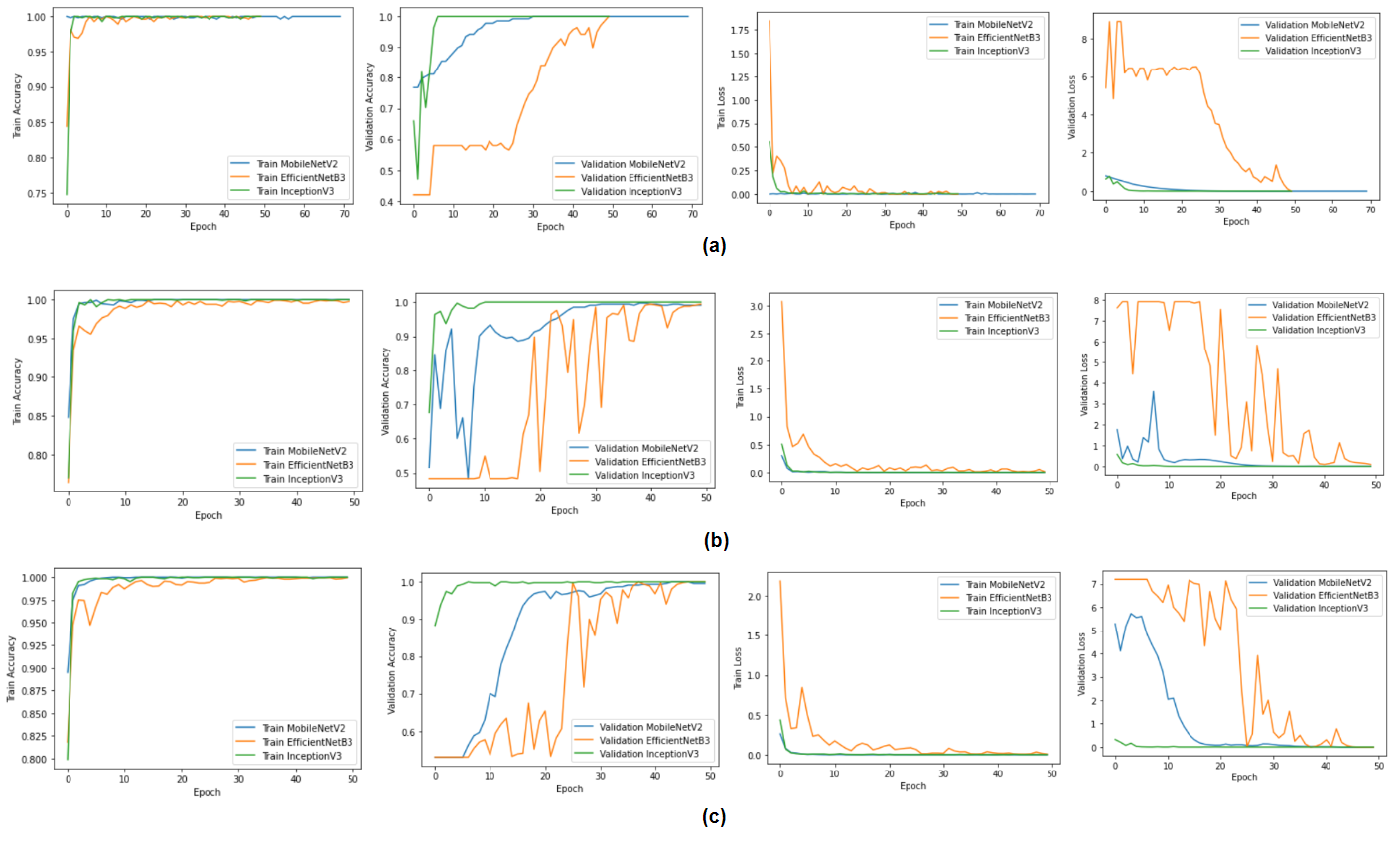} \hspace{0.5cm}
    \caption{Accuracy-Epoch vs Loss-Epoch curves for MobileNetV2-GRU, EfficientNetB3-GRU and InceptionV3-GRU methods (a) the first dataset, (b) second dataset and (c)combined dataset. }
    \label{fig:ap}
\end{figure}

Loss-epoch \& accuracy-epoch curves (MobileNetV2-GRU, EfficientNetB3- GRU and InceptionV3-GRU methods) for the first, second and combined
datasets are presented in Figure \ref{fig:ap}. InceptionV3-GRU model accuracy loss curves are perfectly align for three datasets. As the iteration times increase, the MobileNetV2-GRU model
significantly outperforms compared to the
 EfficientNetB3-GRU model.

\section{Discussion}
Finally, we have performed  CNN models like  EfficientNetB3, InceptionV3 and
MobileNetV2 on the same dataset but got the best classification rate from
our proposed model. For ALL-IDB1 dataset we got 54.54\% for EfficientNetB3, 61.36\% for MobileNetV2 and 95.45\% accuracy  for  InceptionV3 respectively. 69.23\% for EfficientNetB3, 86.53\% for MobileNetV2 and 92.30\% for  InceptionV3 accuracy for For ALL-IDB2 dataset and for combine dataset we got 72.97\% for EfficientNetB3, 94.89\% MobileNetV2 and 89.47\% accuracy  for  InceptionV3 model respectively. The results of our proposed  model are compared with these models that are shown in
the Tables \ref{tab:discussion}.
\begin{table}[ht]
\caption{Comparison of Proposed Model with Individual Model.}
\centering

\begin{tabular}{p{3cm}p{3cm}p{3cm}} 
\hline
\multicolumn{1}{l}{Dataset} & \multicolumn{1}{l}{Methodology}      & \multicolumn{1}{l}{Accuracy} \\ \hline
\multirow{5}{*}{ALL-IDB1 }        & EfficientNetB3   & 54.54\%       \\ 
       & MobileNetV2 & 86.36\%     \\ 
           
     & InceptionV3    & 95.45\%       \\ 
     & Proposed Methodology        & 100\%
     
          \\ \hline
\multirow{5}{*}{ALL-IDB2 }         & EfficientNetB3 & 69.23\%     \\ 
      & MobileNetV2 & 86.53\%    \\ 
    & InceptionV3 & 92.30\%    \\
    & Proposed Methodology        & 98.08\%
    
    \\ \hline
\multirow{5}{*}{ALL-IDB1 +}        & EfficientNetB3     & 72.97\%    \\ 
\multirow{5}{*}{ALL-IDB2}        &     &    \\ 
 & MobileNetV2  & 94.59\%       \\
 & InceptionV3  & 89.47\%       \\
           & Proposed Methodology        & 98.64\% 
           
           \\ \hline  
\end{tabular}

\label{tab:discussion}    
\end{table}

The outcomes of the suggested model have also been compared to the results of research already done on the identification of leukemia malignancy. There have
been a number of methods proposed that have yielded positive results. The proposed work’s accuracy measure is
compared to those of works using the same dataset as shown in Table \ref{tab:dis}. 

\begin{table}[ht]
\caption{Comparison of Proposed Model with Related Study.}
\centering

\begin{tabular}{p{3.5cm}p{3cm}p{2cm}p{1cm}} 
\hline
\multicolumn{1}{l}{Authors} & \multicolumn{1}{l}{Methodology} & \multicolumn{1}{l}{Dataset}      & \multicolumn{1}{l}{Accuracy} \\ \hline
Jyoti Rawat \textit{et al.} \cite{rawat2017classification}        & SVM   & ALL-IDB   & 89.0\%    \\ 
   Sarmad Shafique \textit{et al.} \cite{shafique2018acute}        & AlexNet   & ALL-IDB   & 96.06\%     \\ 
  Nizar Ahmed \textit{et al.} \cite{ahmed2019identification}        & CNN   & ALL-IDB   & 88.25\% \\
    T. T. P. Thanh \textit{et al.} \cite{thanh2018leukemia}       
     & CNN    & ALL-IDB & 96.6\%       \\ 
     Maryam Bukhari \textit{et al.} \cite{bukhari2022deep} & Squeeze and excitation based CNN   & ALL-IDB   & 98.3\% \\
     & Proposed Methodology        & ALL-IDB1 & 100\% \\
     & & ALL-IDB2 & 98.08\% \\
     & & ALL-IDB1+ & 98.64\%
     
          \\
           & & ALL-IDB2 & 
     
          \\\hline

\end{tabular}

\label{tab:dis}    
\end{table}

Earlier, several research studies suggested procedures  optimize classification accuracy. Some of these techniques used segmentation which divides each leukocyte cell into its nucleus and cytoplasm; feature extraction; feature dimensionality reduction which maps the
higher feature space to the lower feature space using principal component analysis; and classification which makes use of common classifiers including k-nearest neighbor, probabilistic neural networks, adaptive neuro fuzzy inference systems, support vector machines, and smooth support vector machines \cite{rawat2017classification} achieved accuracy 89\% and A technique based on convolutional neural networks is used to identify normal and infected blood cells by extracting features from raw blood cell images and classifying them \cite{thanh2018leukemia} with accuracy 96.6\%, while others utilize pretrained AlexNet which was fine-
tuned. Data augmentation was utilized to verify the performance over different color images and reduce overtraining by comparing data sets with different color models\cite{shafique2018acute} accuracy 96.06\%. In \cite{ahmed2019identification} Convolutional Neural Network based method used for classification and obtain accuracy 88.25\%.Squeeze and excitation based CNN is used in \cite{bukhari2022deep} and their proposed method achieves an accuracy of 98.3\%. In \cite{anilkumar2021automated} AlexNet, VGG16, VGG-19, Inceptionv3, MobileNet-v2, Xception, DenseNet-201, Inception-ResNet-v2, ResNet-18, ResNet-50, and ResNet-101 are pretrained series networks that are used for classification. With all of the pre-trained networks utilized in the ALL-IDB1 study, a classification accuracy of 100\% is achieved. All of the techniques employ  enhance classiﬁcation accuracy but none of them  consider the 
uncertainty of the model’s output.
Our model also has the advantage of using numerous models rather than just one during the classification step. In other words, rather than having a single decision process, our model has a setof them.

\section{Conclusion}
\label{conclusion}
One of the main causes of cancer-related mortality is leukemia. Recent research studies that have  been conducted suggest using deep learning techniques, such as transfer learning algorithms, to diagnose leukemia malignancy with remarkably precise outcomes. However, enhancing deep learning algorithms remains an ongoing research challenge among various researchers. The main goal of our work, an improved deep learning model to introduce a new state-of-art deep learning model, but assess the performance of uncertainty quantification to improve the performance of computer-aided diagnostic systems. The proposed study described the use of DL networks for automated leukemia identification with the ALL-IDB, a publicly available dataset. Strong, pertinent, and discriminative features are extracted from leukemic and normal cells using GRU, which enables each model to reduce the gradient vanishing problem. The research can distinguish between normal and leukemia-related images in the dataset with 100\% classification accuracy for the ALL-IDB1 dataset, 98.07\% for ALL-IDB2 dataset and 98.64\% for combine datasets.These methods have demonstrated outstanding results, enabling hematologists to diagnose acute leukemia at an early stage. Leukemia is detected in general by the image classifications employed in the study; differentiation of leukemia into distinct types and subtypes is not taken into consideration. Future research can explore the prospect of classifying leukemia into many types and subtypes by experimenting with different datasets that include images of various forms of the disease.

\section*{Declarations}
The authors do not have any conflicts of interest to declare in 
connection with the publication of this paper. 

    
    



\bibliography{sn-bibliography}


\end{document}